\documentclass[prl,aps,twocolumn,showpacs]{revtex4}
\usepackage{epsfig} % include EPS-figures
\usepackage{graphics}

\pdfoutput=1

\newcommand*{\eps}{\varepsilon}

\begin{document}
%\title{Molecular-Dynamics simulations of crystallization in the microsecond time scale}
\title{Crystallization of the Wahnstr\"{o}m Binary Lennard-Jones Liquid}
\author{Ulf R. Pedersen, Nicholas P. Bailey, Jeppe C. Dyre and Thomas B. Schr\o{}der}
\affiliation{DNRF Centre ``Glass and Time'', IMFUFA, Department of Sciences,
Roskilde University, Postbox 260, DK-4000 Roskilde, Denmark}
\date{\today}

\begin{abstract}
We report observation of crystallization of the glass-forming binary Lennard-Jones liquid first used by Wahnstr\"{o}m [G. Wahnstr\"{o}m, Phys. Rev. A {\bf 44}, 3752 (1991)]. Molecular dynamics simulations of the metastable liquid on a timescale of microseconds were performed. The liquid crystallized spontaneously. The crystal structure was identified as MgZn$_2$. Formation of transient crystallites is observed in the liquid. The crystallization is investigate at different temperatures and compositions. At high temperature the rate of crystallite formation is the limiting factor, while at low temperature the limiting factor is growth rate. The melting temperature of the crystal is estimated to be $T_m=0.93$ at $\rho=0.82$. The maximum crystallization rate of the $A_2B$ composition is $T=0.60\pm0.02$.
\end{abstract}
\maketitle

The use of computer models of liquids has played an important role for understanding glass forming-liquids. Computationally undemanding models are attractive, since long simulation times are important. Further more, the model does not have to be specific since the glass transition is universal. A simple model for molecular interactions is the the famous Lennard-Jones pair potential, $U_{ij}(r_{ij})=4\varepsilon((r_{ij}/\sigma_{ij})^{12}-(r_{ij}/\sigma_{ij})^6)$. It is not possible to investigate a single component Lennard-Jones liquid close to the glass transition since the structural relaxation time is of the same order as the crystallization time. This can be avoided by using a binary Lennard-Jones liquid where the liquid consists of two kind of particles with differend radii $\sigma_{ij}$ and binding energy $\varepsilon_{ij}$ \cite{Wahnstrom1991,Kob1994}. In 1991 Wahnstr\"{o}m suggested parameters for a binary Lennard-Jones liquid in order to investigate the metastable undercooled liquid \cite{Wahnstrom1991}. In the following years the model has been used by several others as a standard liquid \cite{Coslovich2006,Lacevic2004,Albano2004,Lacevic2003,Lacevic2003b,Schroder2000,Sastry1999,Fujiwara1996}.% It should be noted that the parameters for a binary Lennard-Jones liquid suggested by Kob and Andersen \cite{Kob1994} is more investigated.

Increasing computer power expands the time scale that can be simulated. On a time scale of microseconds, the Wahnstr\"{o}m liquid is no longer meta-stable, but crystallizes. In this paper we report observations of formations of crystallites and growth. It is possible to investigate a liquid, a highly viscous metastable liquid, a glass and crystallization within the same model. There is a separation of phonons, structural relaxation, formation of crystallites, and crystal growth in time.

The classical understanding of the crystallization process is as follows \cite{Balibar2006}. Crystallization consists of two events: formation of a critical crystallite and subsequent growth. The free energy of a crystallite (below the melting temperature) consists of a positive surface part and a negative volume part. Crystallites below a critical size are unstable since the surface part will dominate. When a crystallite is above the critical size, the volume part will dominate, and the crystallite will grow.

We have performed molecular dynamics simulations of the binary Lennard-Jones liquid suggested by Wahnstr\"{o}m \cite{Wahnstrom1991}. Particles interact via the Lennard-Jones pair potential. The system consists of two types of particles labelled $A$ and $B$. Energy is reported in units of $\eps_{AA}$, length in units of $\sigma_{AA}$, mass in units of $m_{A}$, temperature in units of $T=\eps_{AA}/k_B$, and time in units of $t^*=\sigma_{AA}\sqrt{{m_{AA}}/{\eps_{AA}}}$. Type $B$ particles have a larger radius than type $A$ particles, $\sigma_{BB}=\frac{6}{5}\sigma_{AA}$ and $\sigma_{AB}=\frac{1}{2}(\sigma_{BB}+\sigma_{AA})=1.1\sigma_{AA}$ and twice the mass, $m_{B}=2m_{A}$. All binding energies are equal, $\eps_{AA}=\eps_{BB}=\eps_{AB}$.

We define Argon units such that $A$ particles may be considered to be Argon molecules; $\eps_{AA}=1$ kJ/mol, $\sigma_{AA}=0.34$ nm $m_A=40$ g/mol, $\eps_{AA}/k_B=120.27$ K and $t^*=2.15$ ps.

The simulations were performed using Gromacs software \cite{gromacs}. A switch function from $2.0\sigma_{AA}$ to $3.5\sigma_{AA}$ was used to make potential go smoothly to zero at $3.5\sigma_{AA}$. Periodic boundaries were imposed. The Verlet velocity integrator with a time step of $0.01t^*$ were used. The temperature was held constant using the Nos\'{e}-Hoover thermostat \cite{Nose1985}. The density were $N/V=0.75 \sigma^{-3}_{AA}$ and $N_A=N_B=512$. This is the same density and $N_A/N_B$ ratio as originally used by Wahnstr\"{o}m \cite{Wahnstrom1991}, but twice the size. We also performed simulations with $N_A=2N_B$ and a density of $0.82\sigma^{-3}_{AA}$. The program VMD was used for visualization \cite{VMD}.

 %%%%% Figures

\begin{figure}
\begin{center}
\includegraphics[width=8cm]{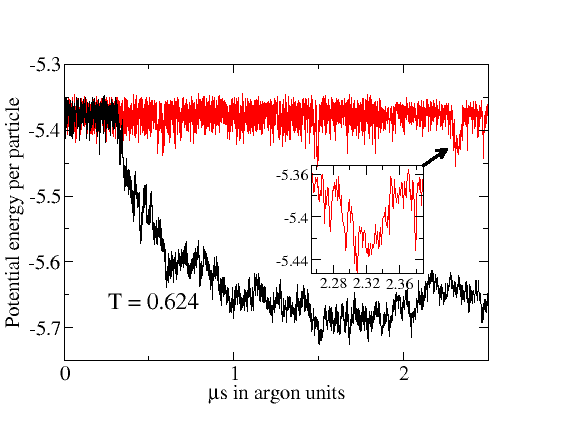}
\includegraphics[width=8cm]{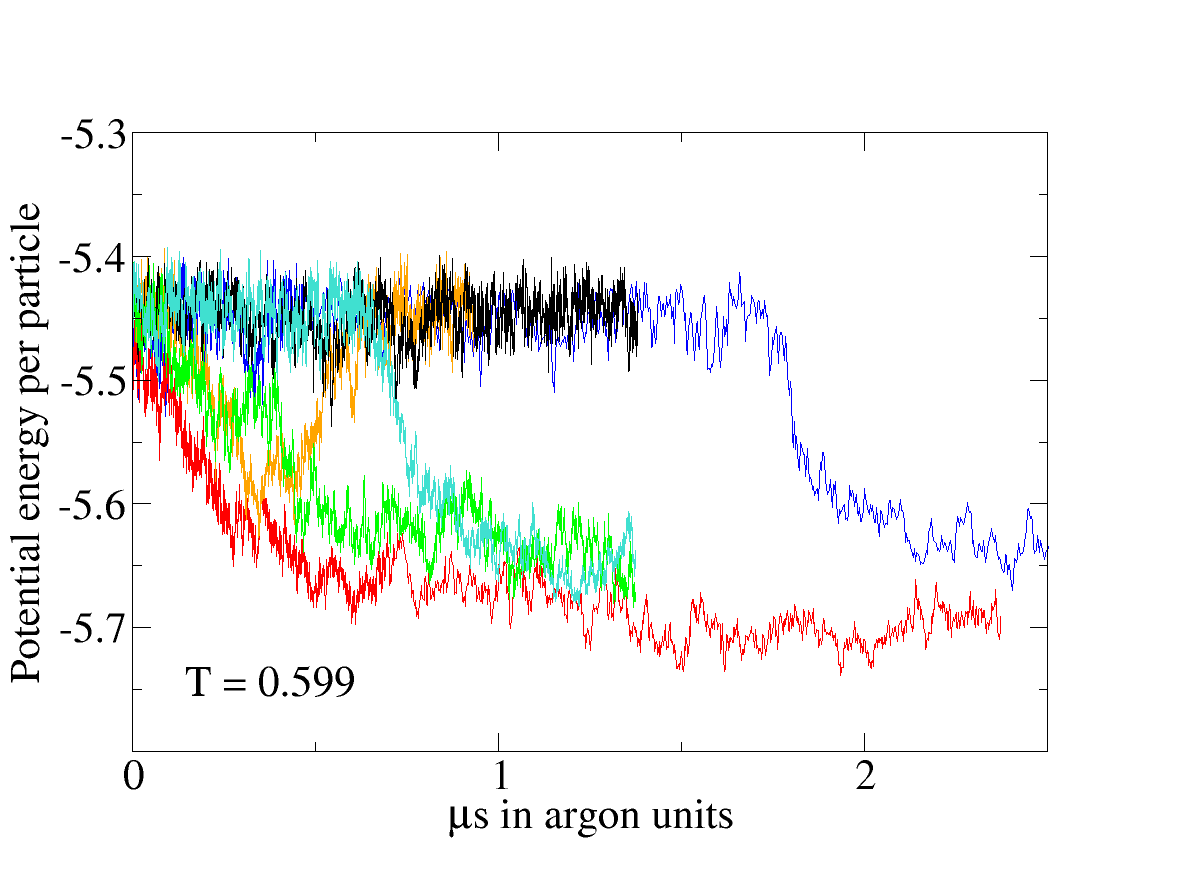}
\includegraphics[width=6cm]{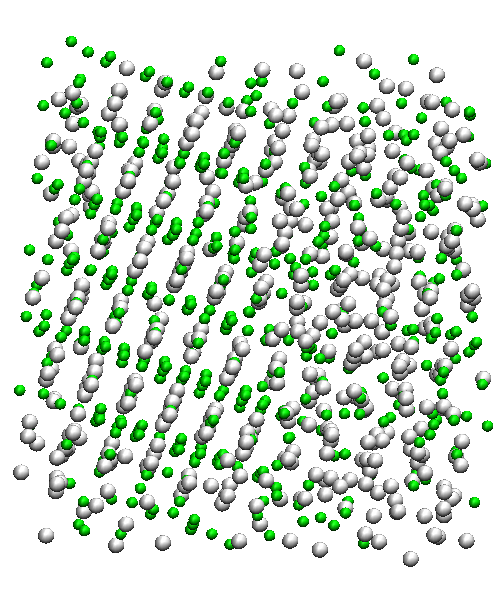}
\caption{The two top panels show the potential energy per particle as a function of time for independent runs at two temperatures with $N_A=N_B=512$. Several of the samples show a dramatic drop in energy due the growth of crystallites. At $T=0.624$ the total simulation time of liquid is $t=13$ $\mu$s (we made nine samples, not all data are shown), but only one sample crystallizes. The inset on the top panel shows a transient drop in energy which can be identified as temporary formation of a crystallite. At $T=0.599$ crystallization happens more often. The bottom panel the quenched structure of the crystallized sample at $T=0.624$. Type A particles are green (dark gray) and type B particles are white (light gray). A crystal has formed with $N_A/N_B\simeq2$, see Fig.\ \ref{CryStr}. The excess $B$ particles and some $A$ particles form a liquid in the remained of the box. It was not possible to crystallize the whole sample within 4.5 $\mu$s.} \label{original}
\end{center}
\end{figure}

 \begin{figure}
 \begin{center}
\includegraphics[width=8cm]{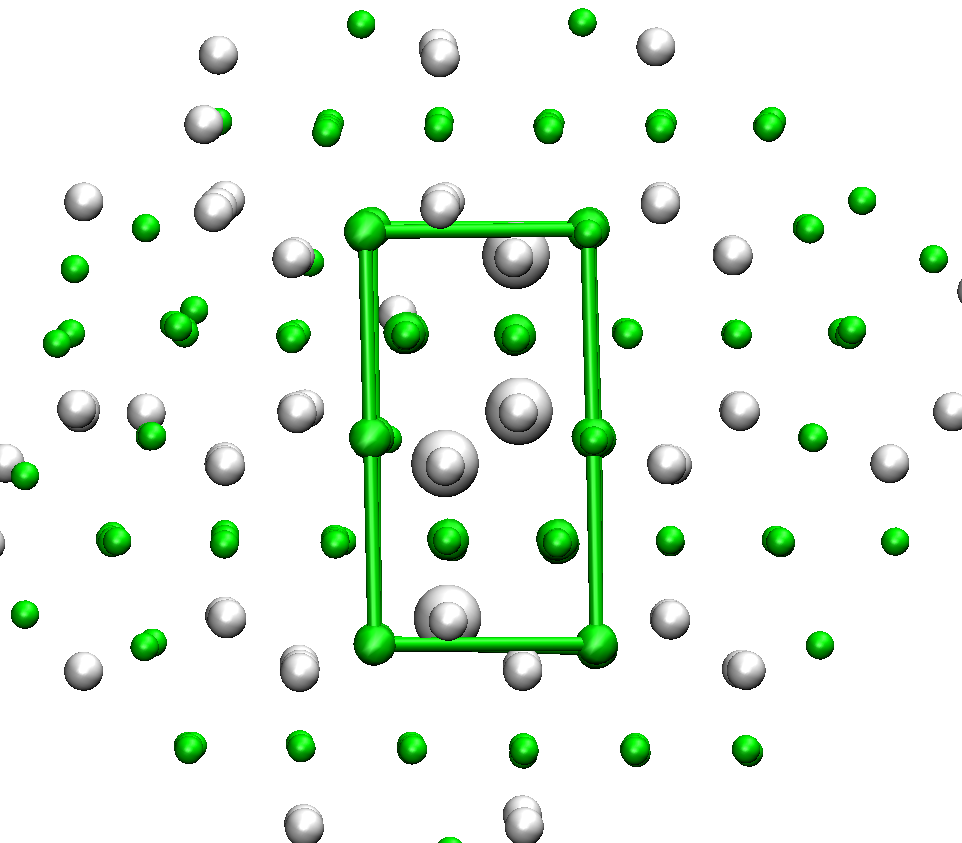}
\includegraphics[width=8cm]{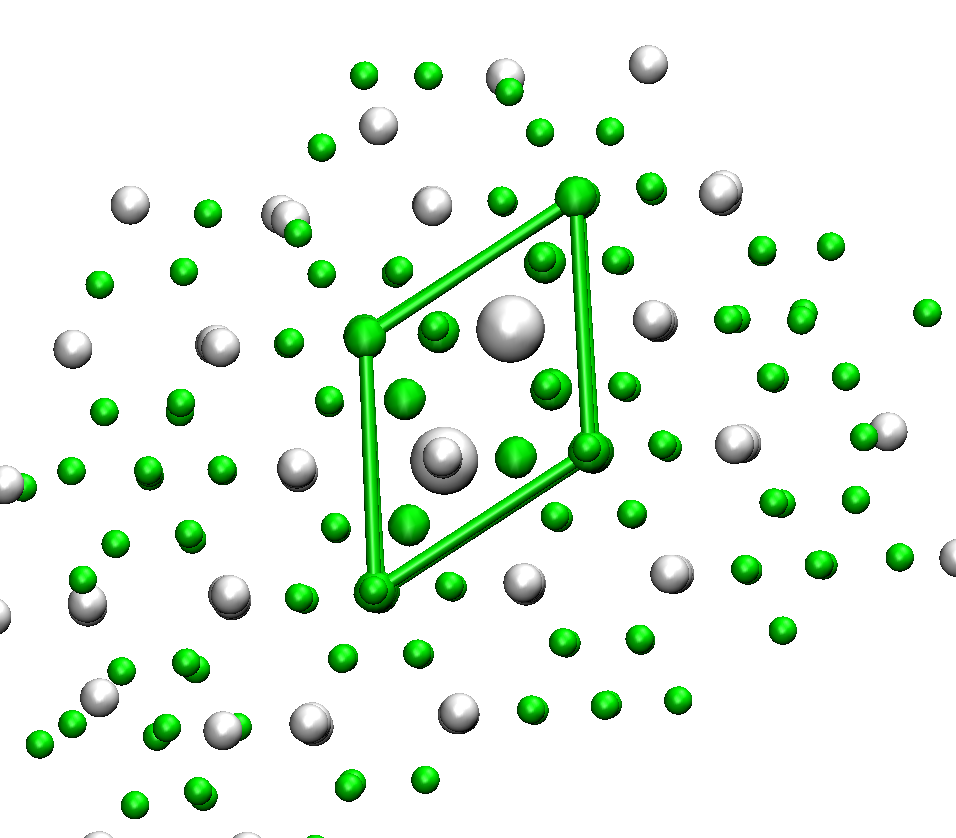}
\caption{Part of crystal seen from different perspectives. The outline of a trigonal, hexagonal unit cell is sketched. Type A particles are colored in green (dark gray) and type B particles in white (light gray). The crystal structure is MgZn$_2$ (C14) with P6$_3$/mmc symmetry. There are 6 type A particles on the \textit{h} position, 2 on the \textit{a} position and 4 type B particles on the \textit{f} position (where \textit{h}, \textit{a} and \textit{f} are Wyckoff letters) \cite{Pearson1972p657}.}
\label{CryStr}
\end{center}
\end{figure}

\begin{figure}
\begin{center}
\includegraphics[width=10cm]{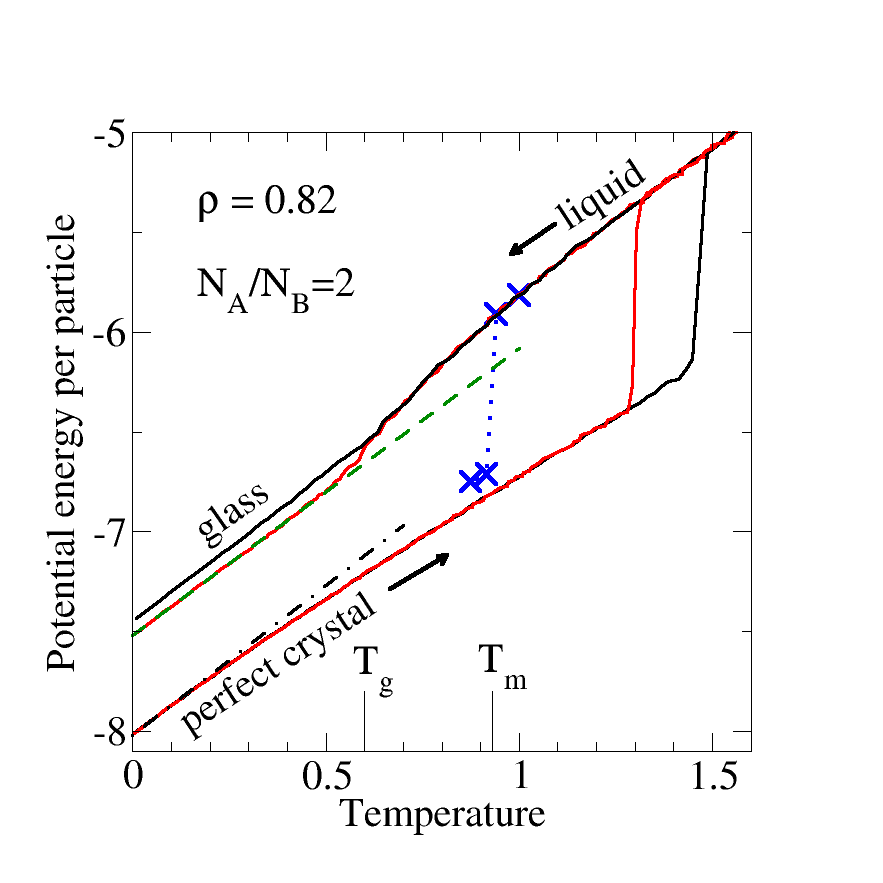}
\caption{The lower line shows the potential energy per particle during heating of a perfect crystal. The heating rates were $2.2\times10^4$ K/$\mu$s and $2.2\times10^2$ K/$\mu$s in Argon units. At $T=1.3$ and $T=1.5$ respectively the crystal melted. The systems were then cooled at the same rates. The liquids undergoes a glass transition at $T_g\simeq0.6$ (depending on cooling rate). The dashed line is a linear fit to the glass data with a slope of 1.44. A partially melted crystal (approximately 90\% crystal) was instantaneously cooled and kept at constant temperature $T'$ for 30 ns. The $\times$'s mark the final energy at different $T'$ temperatures. From this data, the melting temperature was estimated to be $T_m=0.93\pm0.02$. The dash-dot line have a slope of 3/2 corresponding to a classical harmonic crystal.}
\label{dyn1}
\end{center}
\end{figure}

\begin{figure}
\begin{center}
\includegraphics[width=10cm]{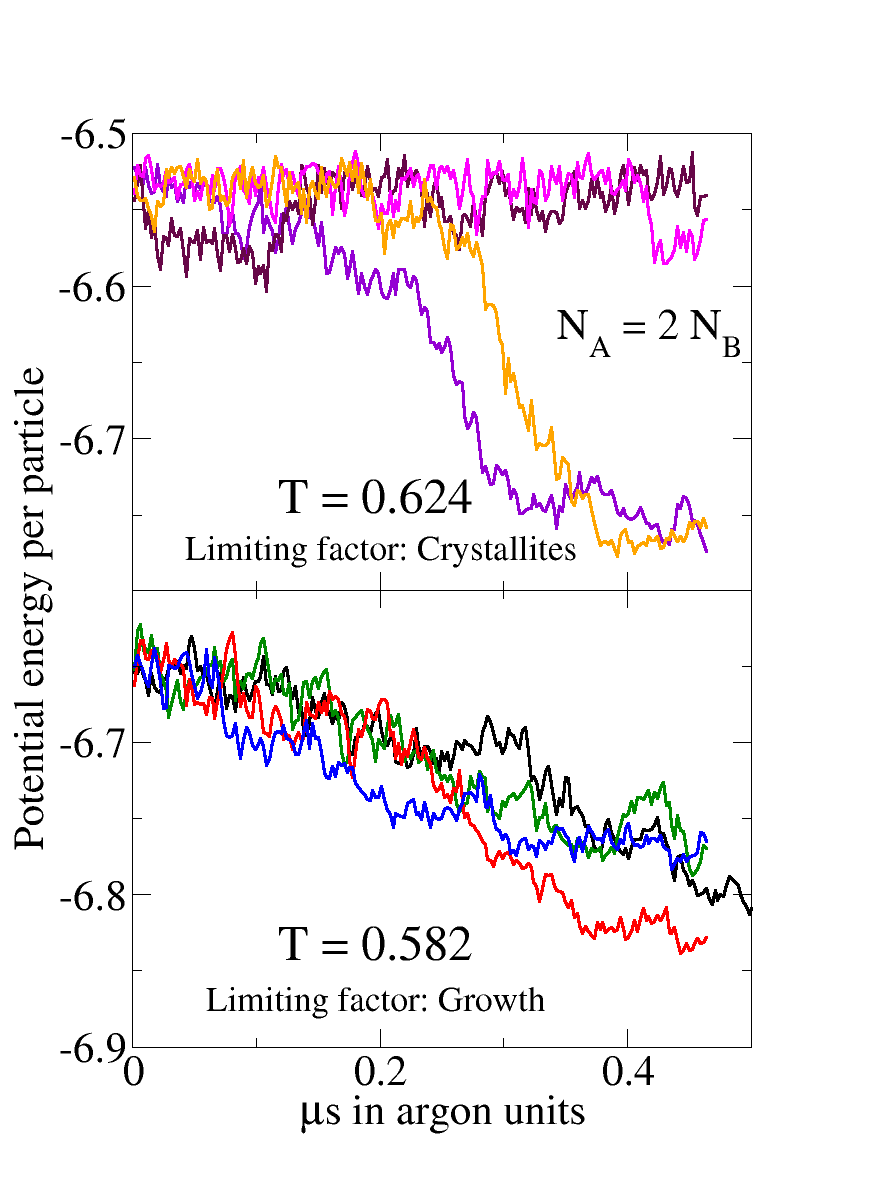}
\caption{Potential energy per particle during simulations of metastable liquids at $T=\{0.625,0.583\}$, $\rho=0.82$, $N_A=1024$ and $N_B=512$. Four independent samples are shown at each temperature. Six of the eight samples crystallizes. At $T=0.625$ the limiting factor is the formation rate of crystallites, while growth rate is the limiting factor at $T=0.583$.}
\label{dyn2}
\end{center}
\end{figure}

%\section{Results}
Fig.\ \ref{original} shows the potential energy per particle as a function of time for independent samples at $T=0.624$ and $T=0.599$. The composition was $AB$. At $T=0.624$ the total simulation time covers 15 $\mu$s or in the order of $10^4\tau_\alpha$ structural relaxation times. One of the samples show a dramatic drop in energy due to crystallization. Crystallization is a rare event at this temperature but more common at the lower temperature. The samples that do not crystallize show some transient drops in energy. This can be explained as formation of small crystallites. This was confirmed by a common neighbor analysis (data not shown) \cite{Honeycutt1987}.

% There is a separation between important time scales in the system. At a temperature of $T=0.624$ we can roughly estimate that if $\tau_{\text{vib}}\sim1$ ps is the characteristic phonon time, $\tau_\alpha\sim$0.5 ns is the structural relaxation time, $\tau_{\text{crystallite}}\sim$30 ns is the typical lifetime of small crystallites, $\tau_{\text{growth}}\sim$200 ns is the time it takes for a critical crystallite to fill the box and $\tau_{\text{critical crystallite}}\sim$ 15 $\mu$s is the time it takes for a critical crystallite to form in the liquid, then
% \begin{equation}
% \tau_{\text{vib}}\ll\tau_\alpha<\tau_{\text{crystallite}}\ll\tau_{\text{growth}}<\tau_{\text{critical crystallite}}.
% \end{equation}

The bottom panel on Fig.\ \ref{original} show the quenched structure of the crystallized sample. A crystal with a few defects has formed. It fills half the box, while the rest of the box is occupied by liquid. The crystal consists of more type $A$ than type $B$ particles so that $N_A/N_B\simeq2$, whereas the liquid mainly consists of type $B$ particles.

Fig.\ \ref{CryStr} shows a part of the crystal seen from different perspectives. A unit cell is sketched. The crystal structure is identified as the hexagonal MgZn$_2$ Laves structure. This is a close packing structure for the $A_2B$ composition with $\sigma_B/\sigma_A={\sqrt{3}}/{\sqrt{2}}\simeq1.22$ \cite{Parthe1961}. This is in good agreement with the parameters used.
% see http://cst-www.nrl.navy.mil/lattice/struk/c14.html

Fig.\ \ref{dyn1} shows simulations of a perfect crystal in an orthorhombic cell with 864 type A and 432 type B particles. The density was $\rho=0.82$. The crystal was heated at constant density from zero temperature with a rate of $\frac{dT}{dt}=4\times10^{-4}$ and $\frac{dT}{dt}=4\times10^{-6}$. At $T=1.3$ and $T=1.5$ respectively the crystal melted. We expect the crystal to be over-heated at these temperatures. The melting temperature $T_m$ was estimated by simulating a partially melted sample (approximately 90\% of the volume were crystal) at constant temperature $T'$ for $30$ ns. The sample melted when $T'>T_m=0.93\pm0.02$ and partially recrystallized when $T'<T_m$.
%% TODO report all numbers in argon units

The liquid made from the melted crystal was cooled with a rate of $\frac{dT}{dt}=-4\times10^{-4}$ and $\frac{dT}{dt}=-4\times10^{-6}$. At $T_m$ the sample did not crystallize, but became super-cooled. At $T_g\simeq0.6$ the sample passed though a glass transition (slightly dependent on cooling rate).

In order to investigate the crystallization properties at the A$_2$B composition further, we performed simulations of the super-cooled liquid at constant temperature with $N=1536$ and $\rho=0.82$. Fig.\ \ref{dyn2} shows the potential energy per particle as a function of time at $T=0.624$ and $T=0.582$. We used several independent samples. At the high temperature two out of four sample show a dramatic drop in energy. Here, the limiting factor for crystallization is the formation of a critical crystallite. When a critical crystallite have formed, the rest of the sample crystallizes fast. At the low temperature all four samples have formed a critical crystallite at $t\simeq0$ but the growth rate is slow. From simulations at several temperatures the maximum crystallization rate was estimated to $T=0.60\pm0.02$ (data not shown).

%\section{Conclusion}
In summary, we show for the first time that the Wahnstr\"{o}m binary Lennard-Jones liquid undergoes crystallization when the metastable liquid is simulated for a time of $10^4\tau_a$ close to the glass transition. %Here, phonons, structural relaxation, crystallite dynamics and crystallization can be separated. %The crystal structure can be reduced to a orthorhombic unit cell with 16 $A$ type particles and 8 type $B$ particles.
The crystal structure is identified as MgZn$_2$. This crystal is rather complex, but can form even when the composition is $N_A=N_B$. The melting temperature of the crystal (at the A$_2$B composition and $\rho=0.82$) is $T_m=0.93\pm0.02$. The maximum crystallization rate of the sample is $T=0.60\pm0.02$.

\section{Acknowledgments}
The authors would like to thank Henning Osholm S\o{}rensen for help with the identification of the crystal structure.

This work was supported by the Danish National Research Foundation Centre for Viscous Liquid Dynamics ``Glass
and Time``.

\end{document}